\documentclass[12pt,prb,amsmath,amssymb]{revtex4}
\usepackage{amsmath,amssymb,amsfonts}

\newcommand{\Aname}{Alice}
\newcommand{\Bname}{Bob}
\newcommand{\A}{\item[\Aname :]}
\newcommand{\B}{\item[\Bname :]}
\newcommand{\BM}{Bohmian mechanics}
\newcommand{\QM}{quantum mechanics}
\newcommand{\wf}{wave function}
\newcommand{\dd}{\partial}
\newcommand{\RRR}{\mathbb{R}}
\renewcommand{\Im}{\mathrm{Im}}

\begin{document}

\title{Understanding Bohmian mechanics: A dialogue}

\author{Roderich Tumulka}
\email{tumulka@mathematik.uni-muenchen.de}
\affiliation{Dipartimento di Fisica and
INFN sezione di Genova, Universit\`a di Genova,
Via Dodecaneso 33, 16146 Genova, Italy}

\begin{abstract}
This paper is an introduction to the ideas of Bohmian mechanics, an
interpretation of quantum mechanics in which the observer plays no
fundamental role. Bohmian mechanics describes, instead of
probabilities of measurement results, objective microscopic events.
In recent years, Bohmian mechanics has attracted increasing
attention by researchers. The form of a dialogue allows me to
address questions about the Bohmian view that often arise.
\end{abstract}

\maketitle

\section{First Day: Fundamentals}

\begin{itemize}
\A What, exactly, does \BM\ say?

\B It describes the motion of $N$ point particles in the usual
three-space. Every particle $i$ has at every time $t$ some definite
position $Q_i(t) \in \RRR^3$. The motion obeys the first-order
differential equation
\begin{equation}
\label{motion}
\frac{dQ_i}{dt} = \frac{\hbar}{m_i} \Im \frac{\nabla_i \Psi
(Q_1(t),\ldots, Q_N(t),t)} {\Psi(Q_1(t),\ldots, Q_N(t),t)},
\end{equation}
where $\Im$ means the imaginary part, $m_i$ is the mass of
particle
$i$, and $\Psi$ is a time-dependent complex-valued function on the
configuration space $\RRR^{3N}$ that satisfies Schr\"odinger's
equation
\begin{equation}
\label{Schr}
i\hbar\frac{\dd \Psi}{\dd t} = - \sum_{i=1}^N \frac{\hbar^2}{2m_i}
\nabla_i\cdot\nabla_i \Psi + V(q_1,\ldots, q_N) \Psi,
\end{equation}
where $V$ is the potential energy. (We denote the variables on which
$\Psi$ depends by $q$, and the actual positions of the particles by
$Q$.)

\A And this mechanics is intended to replace
nonrelativistic
\QM?

\B Yes. The idea is that \BM\ is the true \QM. The $\Psi$ function
is the very same \wf\ you know from \QM, and the positions of the
particles are the same you would find if you performed a
position measurement in \QM.

\A So the Bohmian answer to ``wave or particle?'' is ``wave and
particle!''

\B Yes.

\A But, it's very different from the usual \QM\ conceptually, isn't
it? Indeed, it's not a quantum theory at all; it's a classical
theory.

\B It is indeed very different from the usual \QM\ conceptually.
Usually, it is assumed that quantum particles don't have
trajectories. \BM\ has in common with classical theories that it
tells us a clear story about what's happening. 
On the other hand, as we will soon see, \BM\ is in perfect
agreement with all probabilistic predictions of \QM. So, you are
mistaken thinking that \BM\ is not a quantum theory; 
remember that its empirical implications agree with \QM\ (whenever
\QM\ is unambiguous), and disagree with Newtonian mechanics. A
corollary of this agreement is that \BM\ is confirmed by
experience. In particular, the mere existence of \BM\ proves that
the usually assumed nonexistence of trajectories cannot be concluded from
experiment.

\A You will have to explain the agreement with the predictions of
\QM. But, first I have some questions on the dynamics.
Apparently, you have to assume that the \wf\ is not merely
square integrable, but is differentiable.

\B We do assume that the \wf\ is differentiable (except perhaps at a
few exceptional configurations).

\A For all times?

\B For all times. For a reasonably large class of potentials
(including Coulomb), there is a dense subspace in the $L^2$ Hilbert
space of \wf s that will be differentiable for all times (with few
exceptional configurations).

\A And the equation of motion is ill-defined for all nodes, that is,
zeros, of the \wf. What if your trajectory $(Q_1(t),\ldots,
Q_N(t))$ runs into a node?

\B It has been proved \cite{bmex} that for almost all initial
configurations (according to the appropriate measure) and for all
\wf s from a suitable class, the equation of motion has a unique
global solution (that is, for all $t$). Hence, with ``probability 
one'' Bohmian trajectories never 
run into the singularities of the velocity
field, that is, the nodes and the points where the \wf\ is not
differentiable.

\A What is this appropriate measure?

\B The natural measure for counting initial configurations (that
is, for talking about the size of a set of initial configurations)
for the equation of motion with \wf\ $\Psi(t=0)$ is
\begin{equation}\label{Mass}
|\Psi(q_1,\ldots, q_N,0)|^2 d^{3N}\! q,
\end{equation}
where $d^{3N}\! q$ is the volume measure on configuration space.
The measure (\ref{Mass}) defines a measure on the set of solution curves
$(Q_1(t),\ldots, Q_N(t))$ of the equation of motion.

\A Why don't we simply count initial conditions by the volume
measure?

\B For every measure on configuration space, the dynamics will transport
its density function
$\rho(q_1,\ldots,q_N,t)$ according to the continuity equation
\begin{equation}
\frac{\dd}{\dd t} \rho + \sum_i \nabla_i \cdot \Big( \rho \,
\frac{\hbar}{m_i} \Im \frac{\nabla_i \Psi}{\Psi} \Big) = 0.
\end{equation}
If we start with the volume
measure, that is,
$\rho=1$, at some time, the measure will cease to be the volume
measure at other times. So, when starting with the volume measure,
you arbitrarily prefer some point in time. Not so with the measure
in Eq.~(\ref{Mass}). The measure $|\Psi(0)|^2 d^{3N}q$ is
transported by the dynamics to the measure $|\Psi(t)|^2 d^{3N}q$.
This can be easily checked by deriving the continuity equation
\begin{equation}
\label{continuity}
\frac{\dd}{\dd t} |\Psi|^2 + \sum_i \nabla_i \cdot \Big(|\Psi|^2 
\frac{\hbar}{m_i} \Im \frac{\nabla_i \Psi}{\Psi} \Big) = 0
\end{equation}
from the Schr\"odinger equation. Eq.~(\ref{continuity}) means that 
the Bohmian velocity $(\hbar/m_i) \Im (\nabla_i
\Psi/\Psi)$ equals $j_i/|\Psi|^2$, where $j_i$ is
the probability current density (for particle $i$) of the
\wf.

\A So, what you're saying is that the only way (for generic $\Psi$) to
define a measure on the set of solution curves $(Q_1(t),\ldots,
Q_N(t))$ without preferring some point in time is by Eq.~(\ref{Mass}).

\B Precisely. 

\A Quantum mechanics says that $|\Psi(q,t)|^2$ is the probability
density of finding the particles at configuration $q$ when measuring the
positions at time $t$. If position measurements simply reveal the
Bohmian positions, the Bohmian positions must be random and
distributed according to $|\Psi|^2 d^{3N}\! q$. 

\B We have to keep in mind that the \wf\ we are talking about is the
wave function of all particles in the universe. When we talk about the
distribution of measured positions, what we are considering is an
ensemble of small subsystems, all within the same universe, and all
having the same subsystem \wf\ $\psi$. It has been shown \cite{DGZ}
that for the overwhelming majority [according to the measure
(\ref{Mass})] of possible initial configurations of the Bohmian
universe, the configurations of these subsystems look as if they are
random and independently $|\psi|^2$-distributed. We may think of the
initial configuration of our universe as being random, but such an
assumption is not needed here (and perhaps wouldn't make much sense,
just as we don't regard the dimension of space as a random
number). For a subsystem with \wf\ $\psi$, we may always assume the
configuration to be random and $|\psi|^2$-distributed. This statement
is called the quantum equilibrium hypothesis. \cite{DGZ}

\A What about the collapse or reduction of the wave function?
Equation~(\ref{Schr}) implies there is no collapse. But, in the standard
version of \QM, the collapse rule is required for the theory to give
the correct results. Doesn't \BM\ need the collapse as well?

\B No, \BM\ doesn't need an additional collapse postulate. To see why,
we have to distinguish again between the \wf\ $\Psi$ of the universe
and the \wf\ $\psi$ of a subsystem. Since the evolution of $\Psi$ is
described by Equation~(\ref{Schr}) at all times, $\Psi$ never
collapses, as you said.  In contrast, the \wf\ $\psi$ of the part of
the universe on which we do an experiment does effectively collapse as
a consequence of Eqs.~(\ref{motion}) and (\ref{Schr}).

\A You mean, you can \emph{derive} the collapse from
Eqs.~(\ref{motion}) and (\ref{Schr})? It is well known that the
collapse is nonunitary and therefore is in conflict with the
Schr\"odinger evolution! 

\B We \emph{can} derive the collapse. You will see. For simplicity, we
consider a ``measurement'' with only two possible outcomes. And, let
us first suppose a special form of the \wf\ of the universe, $\Psi =
\psi \otimes \phi \otimes \Phi$, where $\psi$ is the \wf\ of the
subsystem on which we perform the ``measurement,'' $\phi$ is the \wf\
of the measuring apparatus, and $\Phi$ is that of the rest of the
world. The symbol $\otimes$ denotes the tensor product of functions,
that is, $\Psi(x,y,z) = \psi(x) \, \phi(y) \, \Phi(z)$, where $x,y,z$
are the configurations of subsystem, apparatus, and the rest of the
world, respectively. $\Phi$ will be irrelevant to our discussion, so
we ignore it here.

\A $\Phi$ is irrelevant because, as long as $\Psi$
is a product such as
$\text{(something)} \otimes \Phi$, Eq.~(\ref{motion}) implies
that the motion of the subsystem and apparatus particles is
independent of what's happening outside.

\B Yes. Suppose $\hat{U}$ is the unitary operator that represents the
time evolution of the \wf\ during the ``measurement'' process.

\A Wait a second: why do you always put these quotation marks around
the word ``measurement''? 

\B Because we should not expect that anything is actually being
measured during what is usually called a ``measurement.'' I'll
return to this point later.

\A Hm. Go on. 

\B Suppose $\phi_0$ is the \wf\ of the apparatus before the
measurement, $\phi_1$ is that corresponding to the result 1, and
$\phi_2$ is that corresponding to result 2. If $\psi_1$ is the
eigenfunction corresponding to result 1 and $\psi_2$ the eigenfunction
corresponding to result 2, we must have that
\begin{subequations}
\begin{align}
\hat{U} (\psi_1 \otimes \phi_0) &= \psi_1 \otimes \phi_1, \\
\hat{U} (\psi_2 \otimes \phi_0) &= \psi_2 \otimes \phi_2.
\end{align}
\end{subequations}
Now, if $\psi = c_1 \psi_1 + c_2 \psi_2$ is not an eigenfunction of
the self-adjoint operator (the ``observable'') corresponding to this
``measurement,'' then the linearity of the Schr\"odinger equation
implies that
\begin{equation}
\label{Messung}
\hat{U} (\psi \otimes \phi_0) = c_1 \psi_1 \otimes \phi_1 +
c_2 \psi_2 \otimes \phi_2.
\end{equation}
The wave functions $\phi_1$ and $\phi_2$ will have very disjoint
configurational support, that is, $\phi_1$ and $\phi_2$ are supported
by the sets $S_1$ and $S_2$, respectively, in the configuration space
of the apparatus particles, and these two sets will not only be
disjoint, but very far apart in configuration space, as they are
macroscopically distinct. (The wave function $\phi_1$ will not
strictly be zero outside $S_1$, but will be very close to zero, such
that, say, 99.9\% of $|\phi_1|^2$ will be concentrated in $S_1$;
similarly for $\phi_2$ and $S_2$.)

\A Then, if the result is displayed by the position of a pointer (with
$10^{23}$ particles) on a scale, all configurations in $S_i$ will have
the positions of all pointer particles close to $i$, and so the
elements of $S_1$ and $S_2$ will differ by one length unit in at least
$10^{23}$ variables.

\B Yes. For all practical purposes, it will be impossible to have
any interference between the two wave packets on the right-hand
side of Eq.~(\ref{Messung}), because for interference, the supports
of the two packets have to overlap in configuration space.

\A I see.

\B So far we have discussed only the \wf. Now, in \BM, the
configuration point of subsystem + apparatus will be, thanks to the
quantum equilibrium hypothesis, random and distributed according to $|
c_1 \psi_1 \otimes \phi_1 + c_2 \psi_2 \otimes \phi_2|^2$, which for
disjointness of supports equals $|c_1|^2 |\psi_1|^2 |\phi_1|^2 +
|c_2|^2 |\psi_2|^2 |\phi_2|^2$. Therefore, the configuration point
will reside $\mbox{in the set }\{ \mbox{subsystem configurations}\}
\times S_1 \mbox{ with probability }|c_1|^2$, and $\mbox{in the set
}\{ \mbox{subsystem configurations}\} \times S_2 \mbox{ with
probability }|c_2|^2$. Note that this result coincides with the
probability predictions of \QM. Furthermore, if the configuration
point resides in the first set, the output of the apparatus will
(unambiguously) read 1.

\A And, in this case, where is the collapsed \wf\ of the subsystem
after the measurement?

\B The future motion of the configuration point will depend only on
the first wave packet $c_1 \psi_1 \otimes \phi_1$ because, as you
can see in Eq.~(\ref{motion}), the velocity depends only on the
value of the \wf\ and its derivatives \emph{at the configuration
point} $(Q_1(t),
\ldots, Q_N(t))$, and the two wave packets never meet again.

\A Aha. Furthermore, I recall that product wave functions such as 
$c_1 \psi_1 \otimes \phi_1$ lead to independent motion of subsystem 
and apparatus, and I can read off from Eq.~(\ref{motion}) that
$c_1 \psi_1$ generates the same motion as $\psi_1$ since $c_1$ cancels
in the quotient. Hence, the subsystem behaves as if it had
\wf\ $\psi_1$.

\B Yes.

\A But somehow, I missed the point where the collapse comes about.

\B If $x,y,z$ are again the configuration of the subsystem, the
apparatus and the rest of the world, respectively, and
$X(t),Y(t),Z(t)$ is the solution of Eq.~(\ref{motion}), we call
$\psi_\mathrm{cond}(x,t) = \Psi(x,Y(t),Z(t),t)$ the \emph{conditional
\wf}\ of the subsystem.  As long as there is no interaction between
the subsystem and anything else, the conditional \wf\ obeys a
Schr\"odinger equation, but ceases to do so during interaction. The
conditional \wf\ collapses, but not so the \wf\ of the universe.  And,
in contrast to the orthodox collapse, the collapse of
$\psi_\mathrm{cond}$ takes place objectively, takes a finite amount of
time, and does not depend on an observer's knowledge.

\A What happens to the second wave packet, $c_2 \psi_2 \otimes \phi_2$?

\B It leads an empty life. It evolves according to Schr\"odinger's
equation, but it doesn't influence the configuration.

\A But if $\Psi$ never collapses, it isn't a product $\psi_1 \otimes
\mbox{(something)}$ after the experiment. And, we assumed it is a
product in the beginning of our discussion of the measurement
process. So, how do you treat any further measurement?

\B It isn't necessary to assume $\Psi$ is a product. We might have
allowed a number of empty wave packets somewhere far away in
configuration space. Suppose $\Psi_\perp$ is such a wave packet, so
that $\Psi = \psi \otimes \phi \otimes \Phi + \Psi_{\perp}$ while the
support of $\Psi_{\perp}$ is macroscopically disjoint from that of
$\psi \otimes \phi \otimes \Phi$ (which contains the configuration
point); then, our discussion still applies. In this case $\psi$ is
called the \emph{effective \wf}\ of the subsystem,\cite{DGZ} and
$\psi_1$ is the effective \wf\ of the subsystem after the
``measurement.''

\A If I understand you correctly, the outcome of the measurement in
general depends on the microstate, that is, the configuration and the
\wf, of the measurement apparatus. In particular, it depends on the
details of $\phi$, and these details are subject to thermal
fluctuations.

\B In principle, yes. But, for practically relevant experiments, it
turns out that the configuration of the apparatus and the details of
its \wf\ don't influence the outcome. The origin of the randomness
is the unknown subsystem configuration. But, different
experimental arrangements corresponding to the same self-adjoint
operator may lead to different outcomes for the same $\psi$ and
the same subsystem configuration.

\A So, the outcome can't be predicted given a self-adjoint operator
and the state (configuration, \wf) of the subsystem?

\B In many cases, it can't. That's why ``measurement'' is quite a
misnomer in this context, because it isn't at all a property of the
subsystem that is being ``measured.''

\A According to \BM. But, in other interpretations\ldots

\B At least you don't know in general. Ask yourself how you know that
a different apparatus (``measuring'' the same ``observable'') acting on
the same subsystem wouldn't have given a different ``measurement''
result.

\A I'll have to think about this. In quantum mechanics ``measurement''
is never understood in the sense of simply revealing a preexisting
quantity, but rather of forcing nature to choose a value.

\B All the more reason to regard the word ``measurement'' as a misnomer. 
The word suggests a meaning in the outcomes which in general the
outcomes don't have. Nobody would call throwing a die a
measurement, as the outcome is not a preexisting quantity.

\A What about the famous quantum paradoxes in \BM?

\B They get resolved (see, e.g., Ref.~\onlinecite{Belldensity}).
Since \BM\ describes the motion of objectively existing particles, there
can't be any paradoxes.
\end{itemize}

\section{Second Day: Bohmian Versus Orthodox quantum mechanics}

\begin{itemize}

\A I see that \BM\ is a \emph{possible} explanation of the quantum
world. But, the particle trajectories can't be observed!

\B The word ``observe'' is somewhat ambiguous. Strictly speaking,
in a Bohmian universe, the particle paths actually \emph{can} be
observed. Let's consider, for example, a single particle, in a
double-slit experiment. We finally observe the position of
the arrival of every single particle on the screen and, because the
equation of motion is of first order in time, we can calculate the entire
trajectory from this position. For instance, we can decide whether
the particle passed the left or the right slit, without disturbing
the interference pattern: for symmetry reasons, all particles that
passed the left slit hit the left half of the screen, while those
that passed the right slit hit the right half of the screen.

\A But, your last proposition cannot be tested empirically.

\B It cannot be tested empirically. But, it's common for physical
theories to have implications that cannot be tested empirically.

\A I didn't have in mind that you could ``observe'' the trajectory
by calculating it.

\B Most observations, be it the mass of the sun or the charge of the
electron, are not done directly, but involve calculations. I
understand, of course, that you had in mind detecting the particle's
position every tenth of a second. But, the interaction involved with
this detection would influence the particle's future motion, so we
won't see the trajectory the particle would have followed if its
position hadn't been detected (though what we observe is a Bohmian
trajectory as well). It's well known that detecting the particle at
the slits of a double-slit experiment will make the interference
fringes disappear.

\A Hence, the trajectory cannot be seriously observed, and the equation
of motion cannot be tested directly.

\B Neither can the Schr\"odinger equation as we can't observe \wf s.

\A Why can't we observe \wf s?

\B Assume I prepare an atom with a certain \wf\ and I give it to
you. You can't find out the \wf\ if I don't tell you.

\A I see. This fact follows indeed from the mathematical rules of the
quantum formalism. But if you give me a million atoms with the same
\wf, I can determine the \wf.

\B Yes, but I don't give you a million, I give you a single one.

\A But it's not clear if the \wf\ is something \emph{real}. It may
be rather the description of our knowledge about the particle.

\B Let's consider a gedanken experiment. Suppose a computer chooses a
\wf\ randomly and prepares an atom with this \wf. Then, it prints out
some data defining a pair of orthogonal subspaces of the Hilbert
space, one of them containing the \wf\ it had chosen. And, then it
prints out a note that says \emph{which} of the two subspaces
contained the chosen \wf, puts it into an envelope, and seals
it. After that, the computer erases its knowledge about the \wf. Now,
nobody knows the \wf\ of this atom, and nobody can possibly find
out. But, nature still remembers the \wf\ of this atom, because we
can, according to the rules of the quantum-mechanical formalism, carry
out an experiment that has the two subspaces mentioned earlier as
eigenspaces, break the seal, and compare the prediction with the
actual result. (Strictly speaking, agreement between prediction and
result doesn't imply the \wf\ was contained in one of the subspaces,
but the whole procedure can be repeated, and the computer's prediction
is \emph{always} true.) According to the formalism, the machine can
only accomplish certainty of its predictions if the \wf\ actually lies
in the predicted subspace. So, the \wf\ of the atom is well defined (or
``known to nature'' or ``real'') even in those cases when nobody is
aware of it.

\A Strictly speaking, you gave an example of one case in which the
\wf\ is well defined although nobody knows it. This example doesn't
imply it is always well defined.

\B Strictly speaking, you're right about that. But, it suggests that
\wf s are always well defined, and at least it shows that the \wf\
is not merely a mathematical expression of the observer's
knowledge. And, it shows that there exist things we can't observe.

\A If I understand you properly, what you're emphasizing is we can't
directly check Schr\"odinger's equation by means of (i) measuring
the \wf\ (without disturbing it); (ii) letting it evolve an amount
of time; (iii) measuring the \wf\ again; and (iv) comparing the
result with a numerical extrapolation using Schr\"odinger's
equation.

\B Yes. Isn't that true?

\A Certainly. And, you're saying I shouldn't complain about invisible
trajectories as long as I accept Schr\"odinger's equation.

\B You can put it that way. You can, of course, test both
Eqs.~(\ref{motion}) and (\ref{Schr}) by their more indirect
consequences.

\A But, how do I know the correct description of reality is \BM\
rather than any other interpretation of \QM?

\B There is hardly any other interpretation that is consistent,
accepts the existence of an outside reality, and agrees with the
predictions of the quantum formalism. (For discussions of other
interpretations, see Ref.~\onlinecite{BohmHiley} and
\onlinecite{qtwo}.) In fact, the formalism itself suggests \BM. 
Let me explain how. Recall that the 
formalism states that the \wf\ evolves according
to Schr\"odinger's equation unless we perform a ``formal
measurement.'' Every formal measurement is characterized by a
self-adjoint operator, the possible ``measurement results'' are the
eigenvalues of this operator, the probability of a certain result
is the norm squared \ldots

\A \ldots of the projection of the \wf\ to the corresponding
eigenspace, and this projection is the new \wf\ that remains after
the ``formal measurement.''

\B Note that there is an ambiguity in the formalism because it is not
completely clear which processes are formal measurements.  In
particular, we might either guess the \wf\ of the measurement
apparatus, use Schr\"odinger's equation for calculating the \wf\ of
the composite system (object $+$ apparatus) after the measurement, and
\emph{then} invoke the collapse rule when reading off the pointer
position (or computer printout), or we might guess the self-adjoint
operator corresponding to this apparatus and right away assume a
collapse of the object \wf.

\A It is well known and easy to show that this ambiguity does not
influence the set of possible results nor their probabilities or
probabilities for future formal measurements, and hence the
formalism is unambiguous.

\B In so far as macroscopic predictions are concerned. But, because we
saw that the \wf\ (of the composite system) is well defined in
reality, the question arises: when does the \wf\ collapse in reality?
If you find it difficult to believe that the universe switches off the
natural evolution law for a moment in favor of a different dynamics
collapsing the \wf, then apparently the \wf\ \emph{never}
collapses. In this case, however, the \wf\ of the composite system
will, in general, be a superposition of very different states,
including different laboratory protocols or whatever [cf.\
Eq.~(\ref{Messung})]. In particular, the result is not encoded in this
\wf. Neither is there any randomness appearing.

Therefore, the \wf\ cannot be the complete description of the state
of the composite system. There have to be additional variables that
contain the actual result of the formal measurement. Such variables
often are called ``hidden variables'' because they're not part (or
functionals) of the \wf. But, this name turns out to be a misnomer
if you remember that these variables contain the visible result,
the only thing visible, in fact. Now, the question is, what are
these additional variables? Let's see what the formalism suggests:
the \wf\ is a function of the configuration, that is, of the
particle positions. So, what's simpler than assuming that
``particles'' means particles and that a configuration actually
exists? Indeed, what would be the meaning of the \wf\ being a
function of the particle positions if there were no particle
positions? If we assume that quantum particles have trajectories
too, then the motion of these particles should be guided by the
\wf. The precise formula of Eq.~(\ref{motion}) can be obtained as
the simplest one defining a Galilean invariant theory.
\cite{DGZ}

\A I suppose that whoever says that the orthodox view of \QM\ is wrong
should explain where mistakes were made on the way leading to this
view.

\B The founders of \QM\ were much attracted by the thought that the
words ``momentum,'' ``energy,'' and ``angular momentum'' still have a
meaning in \QM. These words, however, don't have an immediate meaning
(in contrast to ``position,'' which does); their meaning in Newtonian
mechanics comes from the fact that they are conserved
quantities. Without this fact, nobody would be interested in
multiplying mass by velocity. Now, Newtonian mechanics has turned out
wrong, so naively we should expect that these words cease to have a
meaning. But, Heisenberg and others insisted they have a meaning. The
idea was that to define a physical quantity means to specify how to
measure it.\cite{heisenberg} But, this is a dangerous strategy because
you don't know whether your result depends on the details of your
measurement arrangement. There's no problem with defining a quantity
by specifying how to measure it as long as you can predict the
values. Then, you can be sure the value didn't depend on the
arrangement. But, there is a problem as soon as the values are
random. You don't even know you measured anything meaningful, because
whatever definition-in-terms-of-how-to-measure you choose, it will
always produce \emph{some} result.

And, it is interesting which definitions Heisenberg chose: the
definitions he gave were always such that in a Newtonian world, they
would have measured the Newtonian value (of momentum, energy, or
angular momentum, respectively). Isn't that strange? Shouldn't we suspect
that the correct experimental arrangement for measuring momentum (if
such a quantity exists) in a world whose rules differ from Newton's
might differ from that in Newton's world? Insisting on the belief that
Newtonian momentum (energy, angular momentum) measurements reveal
\emph{the} momentum (energy, angular momentum) leads to the orthodox
view of \QM.

\A Is there an ``actual momentum'' in \BM\ like the ``actual
position?''

\B You might define $m\dot{Q}$ as the actual momentum (but it is
not a conserved quantity), or you might define $\langle\psi|
(-i\hbar)\nabla |\psi\rangle$ as the actual momentum (which is a
conserved quantity as long as translation invariance is satisfied).
But, I doubt that such a definition will be helpful for calculations
or for anything, as these quantities need not agree with the outcome
of a ``momentum measurement.''

\A There is a pretty symmetry in \QM\ between position and
momentum. \BM\ destroys that symmetry.

\B There is no such symmetry in \QM. The Hamiltonian breaks it. The
Schr\"odinger equation is a differential equation in the position
representation of the \wf, but it is only a pseudodifferential
equation in momentum representation and just some operator equation in
representations using other bases of Hilbert space.

\A But, you can choose a basis in Hilbert space. That's the symmetry.

\B You may as well Fourier transform Maxwell fields. 
But, that doesn't mean there is a symmetry in classical electrodynamics
between physical (position) space and Fourier space.
\end{itemize}

\section{Third Day: Special Issues}

\begin{itemize}
\A What about spin in \BM?

\B We can replace the Schr\"odinger equation by the Pauli equation
and Eq.~(\ref{motion}) by
\begin{equation}
\frac{dQ_i}{dt} = \frac{\hbar}{m_i} \Im \frac{\sum_s \Psi_s^*
\nabla_i \Psi_s} {\sum_s \Psi_s^* \Psi_s},
\end{equation}
where $s$ is the spin index. It is understood that all functions
($\Psi_s$ and its derivatives) are evaluated at the actual
configuration.

\A So, there is no ``actual spin vector?''

\B No. The spin is rather a property of the \wf.

\A What about identical particles?
The \wf\ has to be antisymmetric for fermions and
symmetric for bosons.

\B OK, let the \wf\ be antisymmetric, respectively, symmetric.

\A Nothing special otherwise? The same equation of motion?

\B Nothing special. The same equation of motion.

\A But, the particles are still labeled by the numbers $1, \ldots, N$
in Eq.~(\ref{motion}), whereas identical particles should not have
such a labeling.

\B For symmetric or antisymmetric \wf s, Eq.~(\ref{motion}) is
invariant under permutations of the labels, so the unphysical
labeling does not affect the trajectories.

\A Something else: The ground state of the hydrogen atom is a
real-valued \wf. So, the Bohmian electron does not move. It stands
still.

\B Yes.

\A That's counterintuitive.

\B Counterintuitive doesn't mean much. It may seem counterintuitive
that, according to Maxwell's theory, the energy in a power cord is
not transported within the wires but within the insulator. For my
part, I don't have too much intuition about the interior of a
hydrogen atom. Perhaps you can explain your intuition to me.

\A Well, the nucleus exerts a Coulomb force on the electron, and
in a stable atom this force should be compensated by some
centrifugal force.

\B So, you mean (Coulomb force) $+$ (centrifugal force) $= 0$? Well,
the centrifugal force is, in general, $-m\ddot{x}$, right? So, your
argument implies $m\ddot{x}=$ (Coulomb force). This relation is
precisely Newtonian mechanics, and we can experimentally test
Newtonian against \BM.  \BM\ wins.

\A But, from \QM\ one expects that if particle paths are to make sense,
they should be Newtonian.

\B The existence of particle paths as such does not imply Newton's
equation. It is a frequent prejudice that particle paths, if there 
are any, should be
Newtonian paths. What you refer to in \QM\ is the fact that if a small
wave packet stays a small wave packet for a time, its (only roughly
defined) ``path'' is more or less Newtonian. But, this path is
something different from the Bohmian particle path (which is always
and precisely defined).

\A OK, I'll give a different example. Suppose a particle is confined
between two impermeable walls. Its \wf\ is a multiple of $e^{ikx} +
e^{-ikx}$, where $k$ is chosen so that the \wf\ vanishes at
the walls. Again, the Bohmian particle stands still.

\B Yes.

\A But, \QM\ says the momentum is, up to small corrections, either
$\hbar k$ or $-\hbar k$, so the particle can't be at rest.

\B The word ``momentum'' doesn't have a meaning.

\A But, we can measure the momentum.

\B Tell me how you measure the momentum.

\A Take away the walls and let the particle move freely for an amount
of time. Then, detect its position. If the amount of time was large
enough and the distance between the walls small enough, we know quite
precisely how far the particle traveled.  Now, divide by time and
multiply by mass.

\B The result of this experiment is perfectly predicted by \BM. The
trajectory of the Bohmian particle in your experiment looks like this:
it is a smooth curve $t\mapsto X(t)$ which is constant, $X(t)=x_0$,
before the walls are removed and which is asymptotic to the line
$X(t)\approx (\hbar k/m) t\ +$ constant\ if $x_0$ lies right of the
center and asymptotic to the line $X(t)\approx -(\hbar k/m) t\ +$
constant\ if $x_0$ lies left of the center. Each of these two cases
occurs with probability 1/2.

\A So the particle slowly accelerates until it reaches the velocity
$\pm\hbar k/m$?

\B Yes.

\A But, I always imagined the particle going back and forth between the
walls, having velocity either $\hbar k/m$ or $-\hbar k/m$, each half
of the time.

\B That's Newtonian mechanics, and Newtonian mechanics is refuted by
experiment.

\A But, Newtonian mechanics for our experiment makes the true
prediction that the particle will, with a certain fixed velocity, move
either in the $x$ or in the $-x$ direction after the walls have been
removed. So why should we give up Newtonian mechanics in this case?

\B Because it can't cope with other experiments, such as the
double-slit.

\A I have another question. You said the \wf\ is something real. So,
\BM\ says the \wf\ is something like a physical field.

\B If you wish to put it that way, yes.

\A But physical fields are always functions on three-space, not on
configuration space. Probability densities are functions on
configuration space.

\B The Maxwell and the gravity fields are functions on three-space,
but this doesn't mean every physical field is a function on
three-space. I can imagine having fields on configuration space.  Why
not? Indeed, I can simulate a Bohmian universe on a computer (by the
way, it is very unclear how to simulate an orthodox quantum-mechanical
universe on a computer); now, what should, say, intelligent life forms
inhabiting this universe think about physical fields? These beings
would be wrong about their world unless they regard $\Psi$ as a
physical field on configuration space, as that is how I simulate it.

\A Isn't existence of actual particle positions more of a
metaphysical question than a physical one?

\B An ancient astronomer might have said that the positions of the
planets in three-space cannot be observed, and so we should restrict
our theories to describing the motion of the planets on the
two-sphere, against the background of fixed stars. Such a view
would certainly have influenced physics, so it would not have been
merely of metaphysical interest. That's why I can't see why the
existence of trajectories should not be a physical question.

\A But, as there is no way of testing \BM\ against orthodox \QM\
experimentally, how do I know the trajectories exist?

\B ``Is it not clear from the smallness of the scintillation on the
screen that we have to do with a particle?'' (J.\ S.\
Bell,\cite{Bellbook} p.~191).

\A How do you know you have the correct trajectories? How do
you know it won't turn out to be necessary to change the equation of
motion one day?

\B In fact, I don't. But, that's not a tragedy. How do you know
Schr\"odinger's equation is correct?

\A It certainly isn't. It's nonrelativistic.

\item[Both:] But, that's not a tragedy.

\A Why should a physicist deal with philosophical questions?

\B \BM\ is a differential equation. Not philosophy. It's the
orthodox view that introduces a number of cryptic philosophical
pronouncements for explaining away the problems of \QM.

\A There's one big objection against \BM: the majority of physicists
believes in the quantum orthodoxy.

\B A philosopher, engineer, mathematician, or chemist might accept
the authority of the majority of physicists. But, if you are a
physicist yourself, you are in the position to decide for yourself.

\A A final question: How should we scientifically answer
metaphysical questions?

\B The debate on \BM\ rather resembles the debate at the beginning of
the 20th century on the question ``In mechanical terms, what does
entropy precisely mean and what does the second law of thermodynamics
precisely state?''\cite{ehrenfest} than a metaphysical debate. I have
to explain this comparison. Look, every physical theory we know is
more or less ill-defined. Newton's $1/r^2$ force law is ill-defined as
soon as two particles collide, the Lorentz force evaluates the Maxwell
field at a singular point, and there are dozens of other
problems. Some of these problems we may safely ignore, some not. Some
theories are better defined than others. My message is that the usual
\QM\ is ill-defined in such a way that you should be dissatisfied with
it. Now, the question is how to make sense out of the formalism of
\QM. The meaning of entropy was discussed in statistical mechanics a
hundred years ago, and it is the meaning of \QM\ that we are
discussing now. And, \BM\ is the best way to make sense out of \QM. If
you're wondering what does really happen during quantum processes,
\BM\ is the most natural answer.

\end{itemize}

\section{Further Reading}

J.~S.~Bell's collected papers on the foundations of quantum
mechanics\cite{Bellbook} contain many excellent articles on the
essential problem with ordinary, orthodox quantum mechanics, and the
existing possibilities for solving this problem. Bell calls \BM\ the
``de Broglie-Bohm theory.'' Reference~\onlinecite{Belldensity} is a
nice short paper explaining how \BM\ solves a paradox.

Bohm's original papers are of historical interest.\cite{Bohm52} You
should keep in mind, however, that they represent the 1952
state-of-affairs, containing errors about the behavior of the
solutions of Eq.~(\ref{motion}) and speculations that have not led
anywhere. Later in his life, Bohm wrote a book on
\BM\,\cite{BohmHiley} together with B.\ J.~Hiley. In this book, you
will find pictures of Bohmian paths and detailed discussions of
special topics. Another source of historical interest is the Fifth
Solvay Congress of 1927,\cite{solvay} where similar ideas were
proposed by L.~de~Broglie. The history of Bohmian mechanics and its
reception is outlined in Ref.~\onlinecite{contingency}.

A detailed overview of \BM\ can be found in
Ref.~\onlinecite{Stanford}. An overview of the mathematical research
on \BM\ up to 1995 is given in Ref.~\onlinecite{survey}. A
comparison of \BM\ with other attempts at finding out what \QM\
means is made in Ref.~\onlinecite{qtwo}.

Reference~\onlinecite{DGZ}, a long research paper, contains a detailed
analysis of how to justify the quantum equilibrium hypothesis, and
Ref.~\onlinecite{op} discusses various aspects of quantum measurements
from a Bohmian perspective.  For extensions of \BM\ to quantum field
theory, see Ref.~\onlinecite{crlet} and the references therein, and
for a perspective on a relativistic version of \BM, see Chap.\ 12 of
Ref.~\onlinecite{BohmHiley}.

\begin{acknowledgments}
I owe my material, and many nice formulations, to the papers of John
S.~Bell, David Bohm, Detlef D\"urr, Sheldon Goldstein, and Nino
Zangh\`\i. I am particularly grateful to Detlef D\"urr for many
discussions I had with him before writing this dialogue. I also thank
Sheldon Goldstein and Travis Norsen for their helpful comments and
suggestions.
\end{acknowledgments}


\begin{thebibliography}{18}

\bibitem{bmex} K. Berndl, D. D\"urr, S. Goldstein, G. Peruzzi, and
N. Zangh\`\i, ``On the global existence of Bohmian mechanics,''
Commun. Math. Phys. \textbf{173}, 647--673 (1995).

\bibitem{DGZ} D. D\"urr, S. Goldstein, and N. Zangh\`\i, ``Quantum
equilibrium and the origin of absolute uncertainty,'' J.
Stat. Phys. \textbf{67}, 843--907 (1992). arXiv.org/quant-ph/0308039.

\bibitem{Belldensity} J. S. Bell, ``De Broglie-Bohm, delayed-choice
double-slit experiments, and density matrix,'' Int. J. Quantum 
Chem. \textbf{14}, 155--159 (1980). Reprinted in
Ref.~\onlinecite{Bellbook}, p. 111.

\bibitem{BohmHiley} D. Bohm and B. J. Hiley, \textit{The Undivided
Universe: An Ontological Interpretation of Quantum Theory}
(Routledge, London, 1993).

\bibitem{qtwo} S. Goldstein, ``Quantum theory without observers:
Part one,'' Phys. Today \textbf{51(3)}, 42--46 (1998); S.
Goldstein, ``Quantum theory without observers: Part two,'' Phys.
Today \textbf{51(4)}, 38--42 (1998).

\bibitem{heisenberg} W. Heisenberg, ``\"Uber den anschaulichen Inhalt
der quantentheoretischen Kinematik und Mechanik,'' Z. Phys.
\textbf{43}, 172--198 (1927). English translation p.~62--87 in {\it
Quantum Theory and Measurement}, edited by J. A. Wheeler and
W. H. Zurek (Princeton U. P., Princeton, 1983)

\bibitem{Bellbook} J. S. Bell, \textit{Speakable and Unspeakable in
Quantum Mechanics} (Cambridge U. P., Cambridge, 1987).

\bibitem{ehrenfest} P. Ehrenfest and T. Ehrenfest, ``Begriffliche
Grundlagen der statistischen Auf{}fassung in der Mechanik,'' (1911) in
Enzyklop\"adie der Mathematischen Wissenschaften, Vol.  IV-4, Art.\ 32
(Teubner, Leipzig, 1917).

\bibitem{Bohm52} D. Bohm, ``A suggested interpretation of the
quantum theory in terms of ``hidden'' variables. I,'' Phys. Rev.
{\bf 85}, 166--179 (1952); D. Bohm, ``A suggested interpretation of
the quantum theory in terms of ``hidden'' variables. II,'' Phys.
Rev. \textbf{85}, 180--193 (1952).

\bibitem{solvay} \textit{Electrons and Photons: The Proceedings of the
Fifth Solvay Congress (1927)}, translated and edited by
G.~Bacciagaluppi and A.~Valentini (Cambridge University Press, in
preparation).

\bibitem{contingency} J. T. Cushing, \textit{Quantum Mechanics:
Historical Contingency and the Copenhagen Hegemony} (University of
Chicago Press, Chicago, 1994).

\bibitem{Stanford} S. Goldstein, ``Bohmian mechanics,'' in {\it
Stan\-ford Ency\-clo\-pedia of Phi\-los\-ophy}, edited by E. N. Zalta,
published online by Stanford University {\tt
{<}http://plato.stanford.edu/archives/win2002/entries/qm-bohm/{>}}.

\bibitem{survey} K. Berndl, M. Daumer, D. D\"urr, S. Goldstein, and
N. Zangh\`\i, ``A survey of Bohmian mechanics,''  Nuovo Cimento
Soc. Ital. Fis. \textbf{110}B, 737--750 (1995).

\bibitem{op} D. D\"urr, S. Goldstein, and N. Zangh\`\i, ``Quantum
equilibrium and the role of operators as observables in quantum
theory,'' J. Stat. Phys. \textbf{116}, 959--1055
(2004). arXiv.org/quant-ph/0308038.

\bibitem{crlet} D. D\"urr, S. Goldstein, R. Tumulka, and N.
Zangh{\`\i}, ``Bohmian mechanics and quantum field theory,''
Phys. Rev. Lett. (in press).  arXiv.org/quant-ph/0303156.

\end{thebibliography}
\end{document}